\documentclass[twoside,onecolumn,9pt]{article}
\usepackage{extsizes}
\usepackage[super,sort&compress,comma]{natbib} 
\usepackage[version=3]{mhchem}
\usepackage[left=1.5cm, right=1.5cm, top=1.785cm, bottom=2.0cm]{geometry}
\usepackage{balance}
\usepackage{mathptmx}
\usepackage{sectsty}
\usepackage{graphicx} 
\usepackage{lastpage}
\usepackage[format=plain,justification=justified,singlelinecheck=false,font={stretch=1.125,small,sf},labelfont=bf,labelsep=space]{caption}
\usepackage{float}
\usepackage{fancyhdr}
\usepackage{fnpos}
\usepackage[english]{babel}
\addto{\captionsenglish}{%
  
}
\usepackage{array}
\usepackage{droidsans}
\usepackage{charter}
\usepackage[T1]{fontenc}
\usepackage[usenames,dvipsnames]{xcolor}
\usepackage{setspace}
\usepackage[compact]{titlesec}
\usepackage{hyperref}

\usepackage{ulem}
\usepackage{soul}

\usepackage{epstopdf}

\definecolor{cream}{RGB}{222,217,201}

\begin{document}

\pagestyle{fancy}
\thispagestyle{plain}
\fancypagestyle{plain}{
\renewcommand{\headrulewidth}{0pt}
}

\makeFNbottom
\makeatletter
\renewcommand\LARGE{\@setfontsize\LARGE{15pt}{17}}
\renewcommand\Large{\@setfontsize\Large{12pt}{14}}
\renewcommand\large{\@setfontsize\large{10pt}{12}}
\renewcommand\footnotesize{\@setfontsize\footnotesize{7pt}{10}}
\makeatother

\renewcommand{\thefootnote}{\fnsymbol{footnote}}
\renewcommand\footnoterule{\vspace*{1pt}%
\color{cream}\hrule width 3.5in height 0.4pt \color{black}\vspace*{5pt}} 
\setcounter{secnumdepth}{5}

\makeatletter 
\renewcommand\@biblabel[1]{#1}            
\renewcommand\@makefntext[1]%
{\noindent\makebox[0pt][r]{\@thefnmark\,}#1}
\makeatother 
\renewcommand{\figurename}{\small{Fig.}~}
\sectionfont{\sffamily\Large}
\subsectionfont{\normalsize}
\subsubsectionfont{\bf}
\setstretch{1.125} 
\setlength{\skip\footins}{0.8cm}
\setlength{\footnotesep}{0.25cm}
\setlength{\jot}{10pt}
\titlespacing*{\section}{0pt}{4pt}{4pt}
\titlespacing*{\subsection}{0pt}{15pt}{1pt}

\fancyhead{}
\renewcommand{\headrulewidth}{0pt} 
\renewcommand{\footrulewidth}{0pt}
\setlength{\arrayrulewidth}{1pt}
\setlength{\columnsep}{6.5mm}
\setlength\bibsep{1pt}

\makeatletter 
\newlength{\figrulesep} 
\setlength{\figrulesep}{0.5\textfloatsep} 

\newcommand{\topfigrule}{\vspace*{-1pt}%
\noindent{\color{cream}\rule[-\figrulesep]{\columnwidth}{1.5pt}} }

\newcommand{\botfigrule}{\vspace*{-2pt}%
\noindent{\color{cream}\rule[\figrulesep]{\columnwidth}{1.5pt}} }

\newcommand{\dblfigrule}{\vspace*{-1pt}%
\noindent{\color{cream}\rule[-\figrulesep]{\textwidth}{1.5pt}} }

\makeatother

\twocolumn[
  \begin{@twocolumnfalse}
\vspace{1em}
\sffamily
\begin{tabular}{m{1.5cm} p{13.5cm} }

&\noindent\LARGE{\textbf{Magnetic properties of \{M$_4$\} coordination clusters with different magnetic cores (M$=$Co, Mn).}} \\
\vspace{0.3cm} & \vspace{0.3cm} \\

 & \noindent\large{Simona Achilli,$^{\ast}$\textit{$^{ab}$} Claire Besson,\textit{$^{c}$} Xu He,\textit{$^{d}$}, Pablo Ordej\'on, \textit{$^{d}$}, Carola Meyer\textit{$^{e}$}, Zeila Zanolli\textit{$^{f,b,d}$} } \\

& \noindent\normalsize{We present a joint experimental and theoretical characterization of the magnetic properties of coordination clusters with an antiferromagnetic core of four magnetic ions. Two different compounds are analyzed, with Co and Mn ions in the core. While both molecules are antiferromagnetic, they display different sensitivities to external magnetic field, according to the different strength of the intra-molecular magnetic coupling. 
In particular, the dependence
of the magnetization versus field of the two molecules 
switches with temperatures:
at low temperature the magnetization is smaller in \{Mn$_4$\}, 
while the opposite happens at high temperature. 
Through a detailed analysis of the electronic and magnetic properties of the two compounds we identify a stronger magnetic interaction between the magnetic ions in \{Mn$_4$\} with respect to \{Co$_4$\}. Moreover \{Co$_4$\} displays not negligible spin-orbit related effects that could affect the spin lifetime in future antiferromagnetic spintronic applications. We highlight the necessity to account for these spin-orbit effects for a reliable description of these compounds.
} 
\end{tabular}

 \end{@twocolumnfalse} \vspace{0.6cm}

  ]

\renewcommand*\rmdefault{bch}\normalfont\upshape
\rmfamily
\section*{}
\vspace{-1cm}


\footnotetext{\textit{$^{a}$~Dipartimento di Fisica "Aldo Pontremoli", Universit\'a degli Studi di Milano, Via Celoria 16, Milan, Italy, simona.achilli@unimi.it}}
\footnotetext{\textit{$^{b}$~European Theoretical Spectroscopy Facilities.}}
\footnotetext{\textit{$^{c}$~Department of Chemistry, The George Washington University, Washington DC 20052, USA.}}
\footnotetext{\textit{$^{d}$~Catalan Institute of Nanoscience and Nanotechnology (ICN2), CSIC and BIST, Campus UAB, Bellaterra, 08193 Barcelona, Spain.}}
\footnotetext{\textit{$^{e}$~Department of Physics, Universität Osnabrück, 49076 Osnabrück, Germany.}}
\footnotetext{\textit{$^{f}$~Chemistry Department, Debye Institute for Nanomaterials Science, Condensed Matter and Interfaces, Utrecht University, PO Box 80 000, 3508 TA Utrecht, The Netherlands.}}

\footnotetext{\dag~Supplementary Information (SI) available: Full experimental details, Heisenberg fit of the SQUID data. Spin dynamics calculation. Relaxed structures}




\section{Introduction}

Molecular magnets 
constitute an excellent platform for molecular spintronics and quantum information storage and processing as their properties can be controlled at the nano/micro-scale during fabrication. 
\cite{MMganzhorn, Coronado2020, delBarco2019}
Coordination clusters formed by an inner magnetic core and a surrounding shell of organic ligands 
can be synthesized to control both the magnetic interactions between the ions within the molecule and the coupling between magnetic core and  environment. \cite{Maniaki}
Single-molecule magnets are particularly attractive for spin-dependent quantum transport applications \cite{Bogani} as the spin retains its orientation in the absence of an external magnetic field and applications can leverage on the technology developed for functionalization with nanoparticles.\cite{Zanolli11, Zanolli12} 

In recent years, research has concentrated on molecular magnets with large overall spin generated by ferromagnetic coupling between magnetic centers.\cite{Oshio05, Brooker09, Pedersen14} 
On the other hand, the incorporation of molecular antiferromagnets in spintronic devices \cite{AFMSpin, Bragato} is still a new area of research. Recent  proposals are only theoretical, and concern molecular AFM crystals \cite{Naka19}
or  systems that can hardly be realized experimentally \cite{Fu21}.
%
The expected advantages of antiferromagnetic coordination clusters are the same 
as for antiferromagnetic spintronic devices, i.e.  robustness against perturbation due to magnetic fields,  absence of stray fields, and  capability to generate ultrafast dynamics and large magnetotransport effects. \cite{Baltz}
Antiferromagnetic molecules can be used to functionalize other organic systems, as carbon nanotubes, with the advantage that the current flowing through the tube does not alter the magnetic properties of the molecules \cite{Meyer,besson2021}
and the low spin-orbit coupling allows long spin-flip lengths and spin lifetimes.

The application of molecular magnets in spintronics and quantum technologies would benefit from molecular design aimed at identifying the most suitable combinations of magnetic ions and organic ligands to ensure long spin coherence times, efficient spin injections and tunable transitions between spin states.\cite{Ardavan}
In order to master this kind of applications, it is essential to understand the details of the magnetic interaction between the transition metal core ions and the
subtle dependence of the magnetic properties of the molecule on the molecular structure. \cite{Murrie} For example, 
Kampert et. al \cite{Kampert09} showed that the magnetic properties of a family of \{Mn$_4$\} antiferromagnets with the general formula 
[(RCO$_2$)$_4$Mn$_4$L$_2$] (R=CF$_3$, CH$_3$, Ph, H$_2$L = 2,6-
bis(1-(2-hydroxyphenyl)\-imino\-ethyl)\-pyridine)  
can be tailored by modyfying the bridging carboxylate ligands, leading to a tunable exchange interaction between the magnetic ions. 
In this work we analyze the same molecular cluster with a different perspective by varying the chemical nature of the inner magnetic core.
Through a joint experimental and theoretical characterization we compare the \{Mn$_4$\} acetate complex with its cobalt analogue. While Mn$^{\textrm{II}}$ centers are adequately described 
by the spin magnetic moment (spin-only model), Co$^{\textrm{II}}$ centers in octahedral or pseudo-octahedral environments are characterized by significant orbital moments, leading to spin-orbit coupling (SOC) effects that could reasonably influence applications in more complex devices. Given the importance of magnetic fields in the study of spintronic devices, our experimental investigation mainly focuses on the changes in the magnetic properties of the complexes in an external magnetic field. 
The theoretical analysis, performed through a first-principles approach and a Heisenberg model Hamiltonian, is necessary to interpret the experiments and explain the differences between the two compounds, mostly due to the magnetic features of the magnetic core ions. 
Our computed spin dynamics highlights the key role played by electron correlation in the magnetic behavior of the complexes.

\section{Methods}
\subsection{Theory}
\label{methods}
Theoretical calculations were performed in the Density Functional Theory (DFT) framework, using a pseudopotential description of the core electrons and atomic orbital basis set, as implemented in the SIESTA code. \cite{Soler02, siesta2020} We adopted the local density approximation (LDA) \cite{LDA,CA} for the exchange-correlation energy functional.
A Hubbard correction for Mn and Co was included to account for the strong Coulomb interaction of localized $d$ electrons. We use $U=6$~eV for Mn and $U=4$~eV for Co, according to the literature. \cite{Kampert09, Carter14}
To evaluate the role of spin-orbit coupling, which is relevant in Co, we also performed calculations including spin-orbit correction using the formalism of Ref. \cite{Cuadrado21}.
The current version of the SIESTA code does not allow to simultaneously include Hubbard and spin-orbit corrections, thus the two effects are treated separately.
The structure was relaxed with a tolerance on the forces on the atoms equal to 0.03 eV/\AA.
In the LDA+U calculations the finesse of the real-space grid (mesh-cutoff) was set to 400 Ry and 
the smearing of the electronic occupation (electronic temperature) 
to 100 K. In order to increase the accuracy in the convergence, SOC calculations where performed with 600 Ry mesh cutoff and 1K electronic temperature.
The structural relaxation has been refined with SOC, starting from the LDA+U equilibrium geometry.
The exchange coupling parameters $J_{i,j}$ were obtained by considering the lowest energy spin configurations of the Mn and Co centers and solving a system of equations (Heisenberg model) in the DFT energies with four $J_{i,j}$ parameters. 

The geometry of the various spin configurations was kept fixed to the ground state one in order to exclusively account for the effect of the spin-flip on the total energy of the molecules.\cite{Zanolli18} 
Further, we exploited the model Heisenberg Hamiltonian
$H=\sum_{i,j}\vec {\hat {S_{i}}}\cdot J_{i,j }\cdot\vec {\hat{S_{j}}}+\mu_B g \vec S \cdot \vec B$
, where $\vec{\hat{S_{i}}}$ is the spin vector of atom $i$ and $J$ is the matrix of the exchange parameters,
to fit experimental temperature and field-dependent magnetization data, as allowed by the implementation in the PHI code \cite{phi}.
In the following we label $J_1=J_{1,4}=J_{2,3}$, $J_2=J_{1,3}=J_{2,4}$, $J_3=J_{1,2}$, $J_4=J_{3,4}$, with the atoms numbered as indicated on Figure~\ref{struct}.

\subsection{Experiment}

The complexes [M$_4$L$_2$(OAc)$_4$] (M = Mn, Co, Zn, \{M$_4$\} for short), where H$_2$L = 2,6-bis-(1-(2-hydroxy\-phenyl)imino\-ethyl)pyridine,
HOAc = acetic acid, and M = Mn$^{\textrm{II}}$, Co$^{\textrm{II}}$ or Zn$^{\textrm{II}}$) were synthesized 
by  one  pot  reaction  of 2-aminophenol, diacetylpyridine and manganese, cobalt or zincacetate,
as described by Kampert \emph{et al.} for the manganese complex,\cite{Kampert09}  with some modifications for the cobalt and zinc analogues. Full details of the synthesis methods are given in the ESI.\dag

The two new molecular complexes were characterized via
single crystal diffraction conducted on a SuperNova (Agilent Technologies) diffractometer using Mo K radiation
at 120~K. The crystals were mounted on a Hampton cryoloop with Paratone-N oil to prevent solvent loss.
Thermogravimetric analysis was performed using a Mettler-Toledo TGA/SDTA 851e instrument with a heating
rate of 10~K/min.
Cyclic voltammograms were recorded in dry and deaerated acetonitrile solutions containing tetrabutylammonium
perchlorate (0.1~M) as electrolyte and 3~mM of the analyte, using a SP-150 potentiostat (BioLogic Science
Instruments) controlled by the EC-Lab software and a standard three-electrodes setup including a glassy carbon
working electrode (diameter 3~mm), a platinum wire counter electrode and an Ag/\ce{AgNO3} (0.1~M) reference electrode.
Ferrocene was used as an internal standard.
Magnetometry was performed on a Quantum Design MPMS-5XL SQUID magnetometer. The crystalline samples
were crushed and placed under vacuum for 16 h before the complete removal of solvate molecules was checked by
TGA. The resulting powders were compacted and immobilised into PTFE capsules. All data were corrected for the
contribution of the sample holder (PTFE capsule). Measurements on \ce{\{Zn4\}} were used to determine the diamagnetic susceptibility of this complex $\chi_{dia}$({Zn$_4$})$=-5.3\times 10^{-9}$ m$^3$/mol.
The diamagnetic contribution in \{Co$_4$\} and {Mn$_4$} was then calculated from this value and Pascal's constants\cite{Pascal} for the Zn$^{2+}$, Co$^{2+}$ and Mn$^{2+}$ ions, yielding $\chi_{dia}$({Co$_4$})$=-5.1\times10^{‑9}$m$^3$/mol and $\chi_{dia}$({Mn$_4$})$=-5.2\times10^{-9}$m$^3$/mol, and subtracted from the experimental susceptibility data.

\section{Theoretical and experimental analysis}

\subsection{Synthesis and redox properties}

\begin{figure}[t]
\centering
  \includegraphics[width=0.48\textwidth]{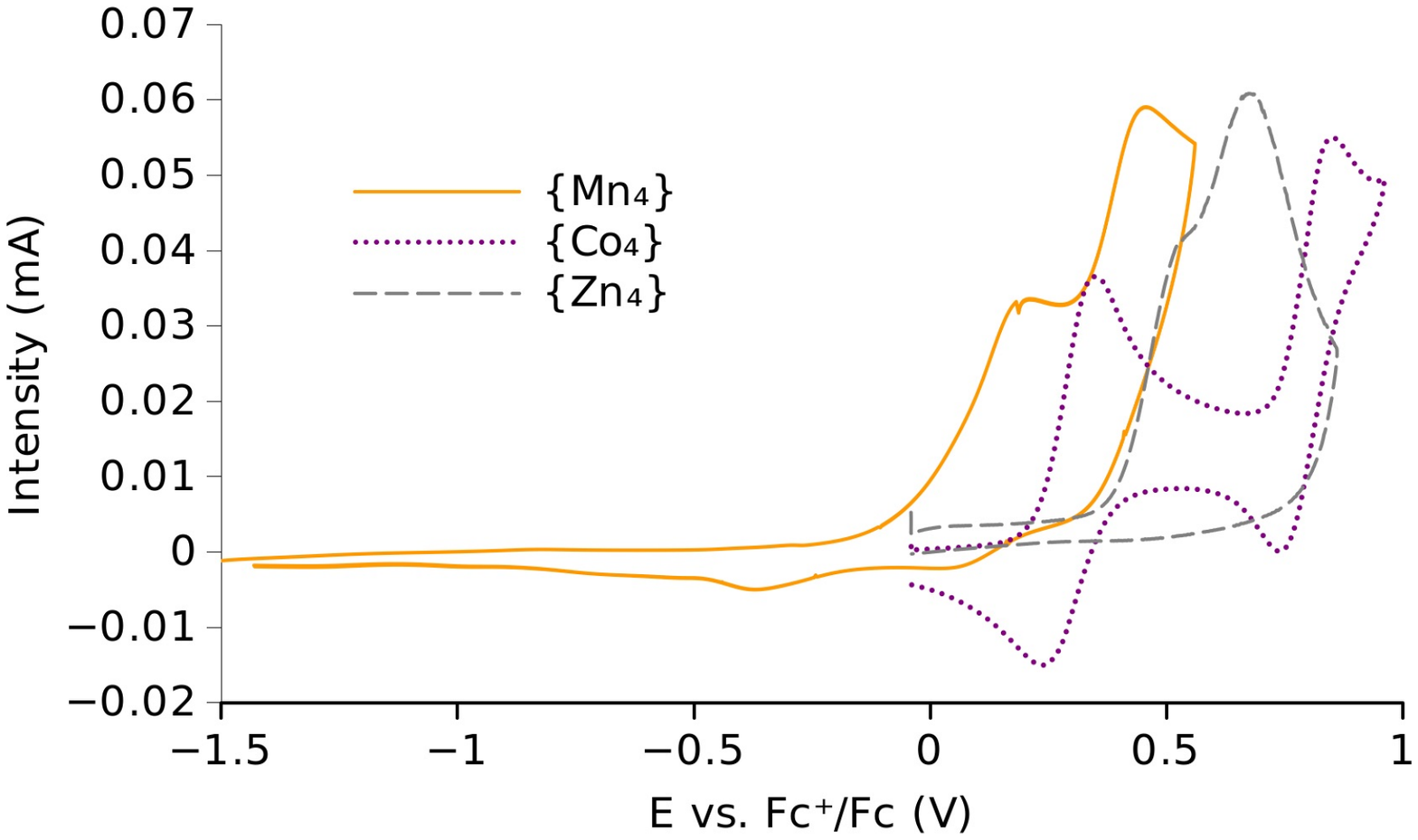}
  \caption{Cyclic voltammograms of the \{M$_4$\} complexes (M = Mn, Co, Zn, concentration ca. 3 mM) in acetonitrile. Tetrabutylammonium perchlorate (0.1 M) is used as electrolyte and the scan rate is 50 mV/s.}
  \label{echem}
\end{figure}

The \{M$_4$\}, synthesized  by one pot reaction of 2-aminophenol, diacetylpyridine and manganese, cobalt or zinc acetate, are stable towards oxidation in the solid state as well as in solution, despite the sensitivity of the \{Co$_4$\} precursors to oxidation by O$_2$ during synthesis. This observation is confirmed by the cyclic voltammetry of the complex (Fig.~\ref{echem}), which displays two quasi-reversible ($\Delta E = 120$~mV) one-electron oxidation waves at 0.30 and 0.80~V vs.~Fc$^+$/Fc which can be assigned as $\{\textrm{Co}^{\textrm{II}}_4\} \longrightarrow \{\textrm{Co}^{\textrm{II}}_3\textrm{Co}^{\textrm{III}} \} \longrightarrow \{\textrm{Co}^{\textrm{II}}_2\textrm{Co}^{\textrm{III}}_2\}$. As expected, the redox couples in the manganese complexes are shifted to lower potentials and show the large peak-to-peak characteristic of $\textrm{Mn}^{\textrm{II}}(\textrm{HS}) \longrightarrow \textrm{Mn}^{\textrm{III}}(\textrm{LS})$ processes. Finally, the zinc derivative shows irreversible ligand-centered oxidation processes above 0.5~V vs.~Fc$^+$/Fc.

\subsection{Molecular structure}
\begin{figure}[t]
\centering
  \includegraphics[width=0.48\textwidth]{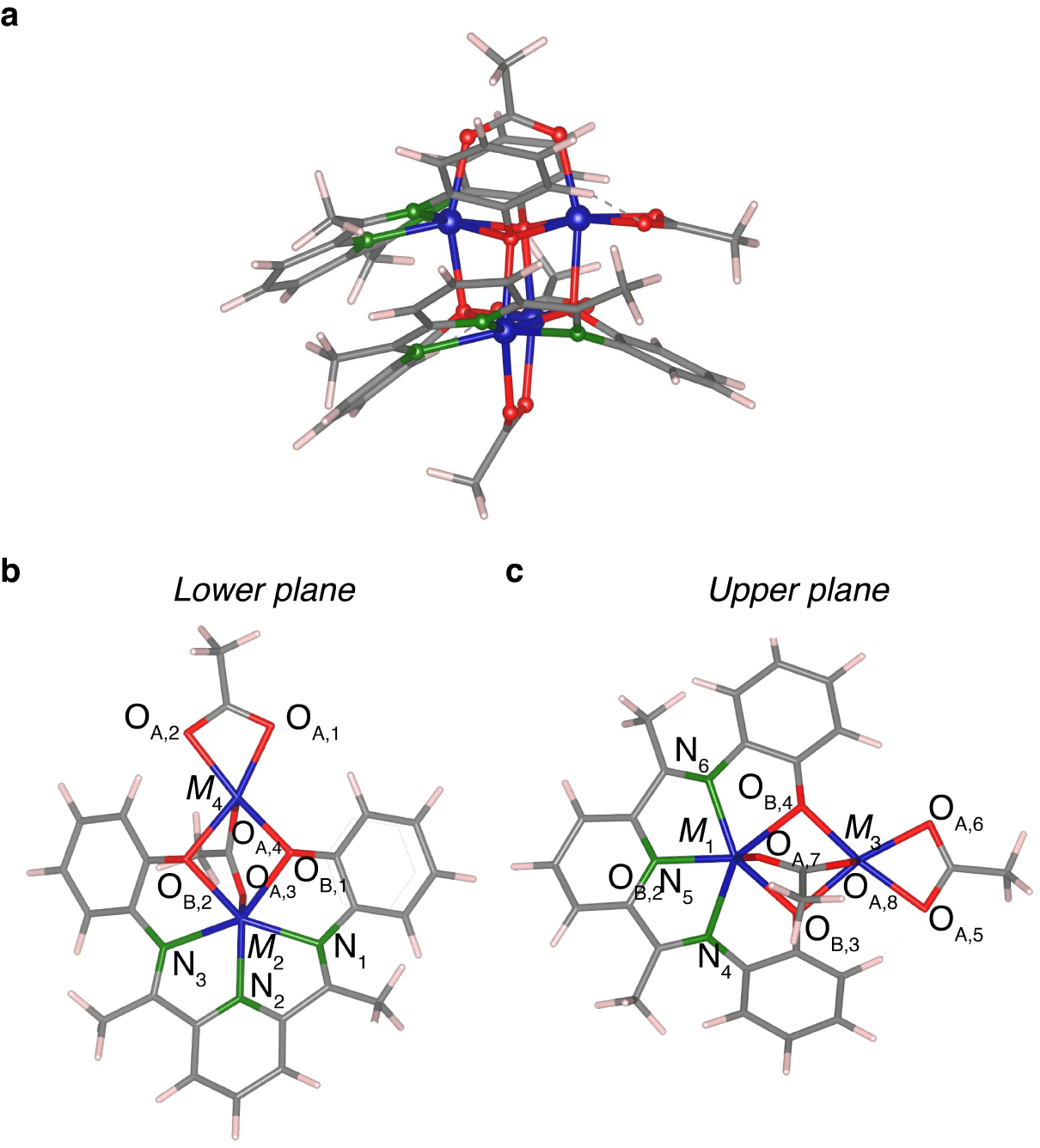}
  \caption{Molecular structure of the {$M_4$} complex. a) side view. Blue: Mn, Co or Zn atoms; red: O; green: N; gray: C, white: H. b) top view of the lower half of the molecule. c) top view of the upper half of the molecule. Relaxed coordinates are available in ESI.}
  \label{struct}
\end{figure}

The structure of the cobalt and zinc complexes was determined by single crystal X-ray diffraction to be analogue to that of the previously published manganese complex\cite{Kampert09}: the complexes consist of a cubic M$_4$O$_4$ core with two sets of two different ligand groups, for a total of 118 atoms (Fig.~\ref{struct}). The metallic core is a quasi-tetrahedron composed of two 7-coordinated ions (M$_1$, M$_2$) with pentagonal bipyramidal coordination and two 6-coordinated ions  (M$_3$, M$_4$) with pseudo-octahedral symmetry.

The metal ions with the same coordination number are almost equivalent, as they overall face a quasi-identical chemical environment. The molecular cluster has an approximately $C_2$ symmetry and can be described as two identical structures on different planes that are rotated of 90$^\circ$ one respect to the other (Fig. \ref{struct}), taking the rotation axis along $z$. 
Seven-coordinated M$_1$ and M$_2$ lie on different planes, 
each being connected in-plane to a pentadentate pyridine-diimine-diphenoxide type ligand ($\textrm{L}^{2-}$), 
completed by the oxygen of a bridging acetate and a $\kappa^3$ phenoxide oxygen from the other $\textrm{L}^{2-}$ ligand.  The coordination sphere of M$_3$, M$_4$ is a pseudo octahedron of six oxygen atoms provided by a bidentate acetate ligand, the other oxygen of the two bridging acetate and two phenoxide oxygen from the $\textrm{L}^{2-}$ ligand.
(Relaxed coordinates available in ESI).

The inner cage of the three molecules is composed by four transition metal ions with different atomic valence configurations, $3d^5$ for Mn(II), $3d^7$ for Co(II) and $3d^{10}$ for Zinc(II). The latter complex is therefore diamagnetic; it was used experimentally to determine the diamagnetic contribution to the susceptibility of the complexes and will not be discussed further. Both Mn(II) and Co(II) ions display high spin configurations, i.e. S=5/2 for Mn and S=3/2 for Co.

Despite the similarity between the structures of the manganese and cobalt complexes, experimental evidence and DFT calculations show small differences in bond lengths in the inner core.
In agreement with the larger atomic radius of Mn with respect to Co, the \{Mn$_4$\} central cage is slightly larger than the \{Co$_4$\} one, due to larger M-O and M-N bond-lengths. Details of the structure are reported in Table \ref{tab-struct}. 

\begin{table}[h]
  \caption{\ Theoretical (DFT) and experimental (XRD) bond-lengths (\AA) of the \{Mn$_4$\} and \{Co$_4$\} molecular complexes. XRD data was obtained at 208~K for \{Mn$_4$\}\cite{Kampert09} and at 100~K for \{Co$_4$\}.}
  \label{tab-struct}
  \begin{tabular*}{0.48\textwidth}{@{\extracolsep{\fill}}rlllll}
    \hline
   d(\AA) & $M_1$-$M_2$  & $M_3$-$M_4$ & $M_{1}$-$M_3^a$& $M_2$-O$_{B,1}^b$  & $M_4$-O$_{A,2}^b$ \\
    \hline
    \{Mn$_4$\}\\
    DFT & 3.35 & 3.45 & 3.64/3.65  & 2.20-2.30 &  2.15-2.22 \\ 
               XRD & 3.62 & 3.47 & 3.50/3.55  & 2.27-2.31 &  2.21-2.22 \\
    \{Co$_4$\} \\
    DFT &  3.26 & 3.21 & 3.42/3.44 & 2.15-2.21 &   2.07-2.11 \\
               XRD &  3.39 & 3.19 & 3.15/3.16 & 2.17-2.29 & 2.11-2.19 \\
    \hline
  \end{tabular*}
  {\it a}: The two values correspond to the equivalent pairs of atoms.\\
  {\it b}: Range given for all equivalent distances in the complex.
\end{table}

\subsection{Behavior in magnetic field}
\begin{figure}[t]
\centering
  \includegraphics[width=0.48\textwidth]{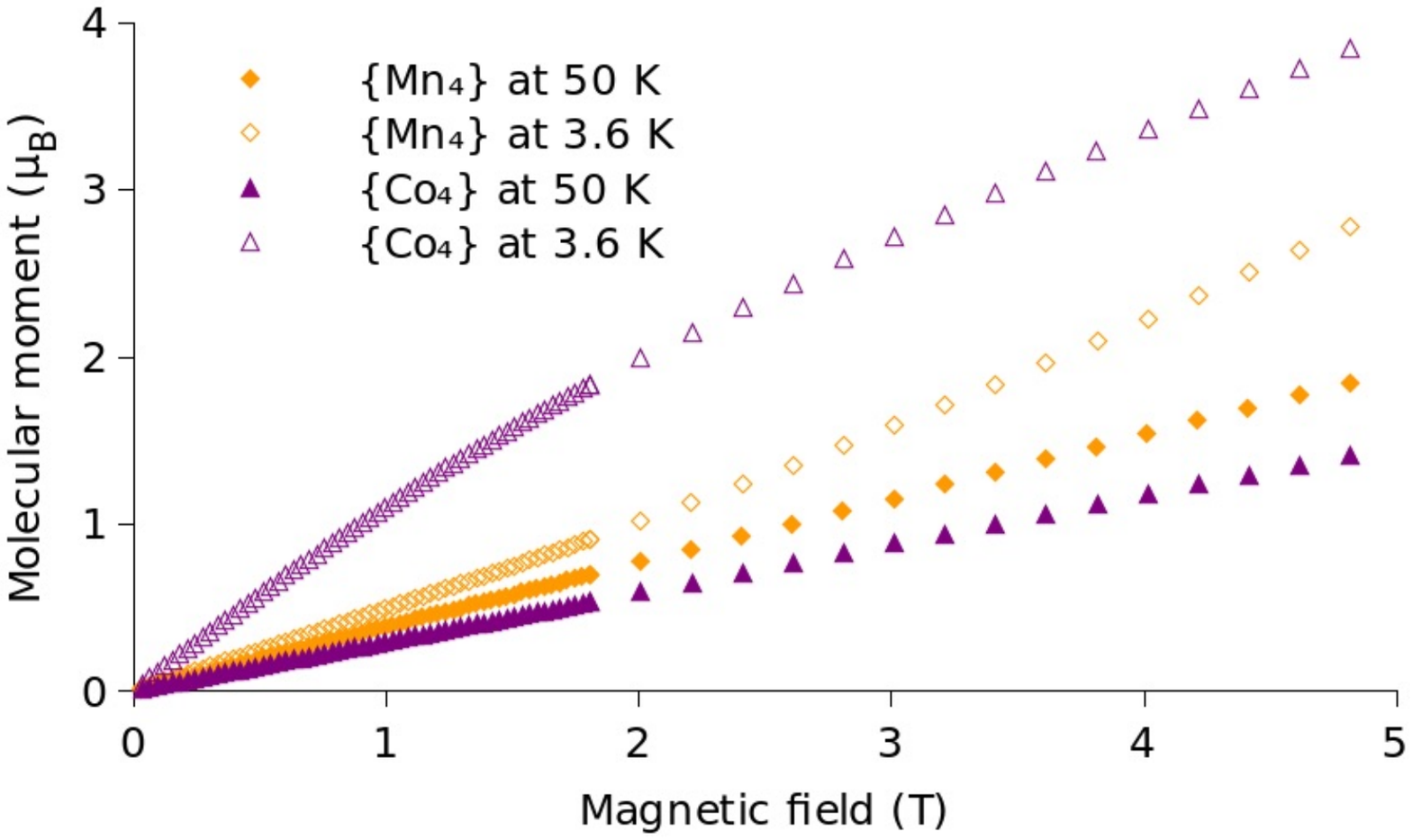}
  \caption{SQUID magnetometry data of \{Mn$_4$\} (orange) and \{Co$_4$\} (purple) at T=3.6 K (open symbols) and T=50 K (filled symbols).}
  \label{SQUID}
\end{figure}

In order to quantify the strength of the magnetic interaction within the molecule and the response to an external magnetic field we performed SQUID magnetometry experiments (Fig. \ref{SQUID}).
A singlet ground state is observed in both complexes, indicating the presence of antiferromagnetic coupling between the magnetic ions in the molecule. The molecular moment $\mu_{mol}$ as a function of magnetic field ($H = 0-5\ \textrm{T})$) and temperature ($T = 20-300\ \textrm{K}$) was fitted to a mean field model (Curie-Weiss law, equation~\ref{CurieWeiss}), yielding N\'eel temperatures of $T_N = 23\ \textrm{K}$ for \{Mn$_4$\} and $T_N = 12\ \textrm{K}$ for \{Co$_4$\}:
\begin{equation}
    \mu_{mol} = \frac{C H}{T-T_N}
    \label{CurieWeiss}
\end{equation}

Those values, as well as the larger slope of the molecular moment at low field/low temperature observed for \{Co$_4$\} in comparison to \{Mn$_4$\}, suggests that the antiferromagnetic coupling between the metal atoms in \{Co$_4$\} is smaller than in \{Mn$_4$\}. 
Notably, the behavior is reversed at high temperature with a larger magnetic moment for \{Mn$_4$\} than for \{Co$_4$\}, which is in agreement with the higher spin moment of the Mn centers.

The Curie constant $C$ obtained for the \{Mn$_4$\} complex ($3.0\times 10^{-3}\ \mu_B\cdot\textrm{K}\cdot\textrm{Oe}^{-1}$) is in good agreement with a spin-only model ($3.1\times 10^{-3}\ \mu_B\cdot\textrm{K}\cdot\textrm{Oe}^{-1}$ for $g=2$ and $S=5/2$). Such a model is not adequate for octahedral cobalt(II) complexes, with their $^{4}T_{1g}$ ground term, and effective orbital momentum $L = 1$.\cite{Pardo2008} Indeed, the Curie constant of \ce{\{Co4\}} ($2.4\times 10^{-3}\ \mu_B\cdot\textrm{K}\cdot\textrm{Oe}^{-1}$) obtained from the Curie-Weiss fit deviates significantly from the one calculated by the spin-only model ($1.3\times 10^{-3}\ \mu_B\cdot\textrm{K}\cdot\textrm{Oe}^{-1}$).

The measured magnetic susceptibility is reported in ESI, together with the theoretical one obtained from spin dynamics calculations. The agreement between theory and experiment is fairly good in the LDA+U approximation (see the discussion in \ref{sec-J}).

\subsection{Magnetic and electronic properties}
\begin{figure}[t]
\centering
  \includegraphics[width=0.48\textwidth]{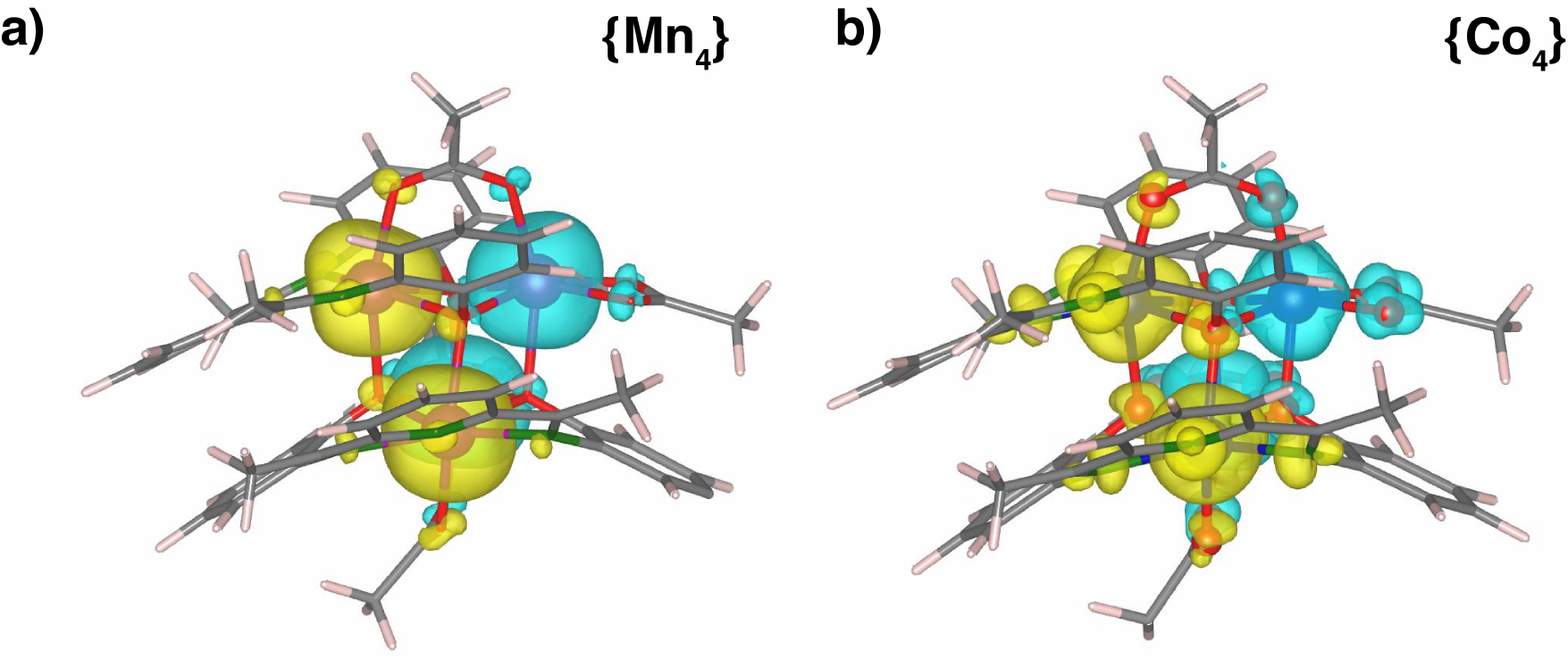}
  \caption{Spin density on \{Mn$_4$\} and \{Co$_4$\}. Yellow (blue) isosurfaces correspond to positive (negative) values with a fixed value of 0.025. Spin density on the ligands is more pronounced in the \{Co$_4$\} case.
  }
  \label{spin-density}
\end{figure}

In order to characterize the magnetic configuration of the inner cage and to ascertain the role of spin-orbit coupling in \{Co$_4$\} we performed ab initio calculations.
DFT analysis shows that the ground state is characterized by an antiferromagnetic coupling between non-equivalent metal ions (M$_1$/M$_3$ and M$_2$/M$_4$), and a ferromagnetic one between the equivalent pairs (M$_1$/M$_2$ and M$_3$/M$_4$) giving rise to a up-up-down-down ($uudd$) configuration, referring to the relative alignment of the spins of the four metal ions.
The magnetic moments of Co, Mn, N and O in the two molecular complexes, deduced from the Mulliken charge population, are reported in Table \ref{tab-spin}.
Due to the chemical interaction with the ligands the magnetic moment of the metal atoms in the molecular complexes is reduced with respect to the isolated ions ($\sim 4\%$ in \{Mn$_4$\}, $\sim 10\%$ \{Co$_4$\}).
Accordingly, the induced magnetization of the ligands is smaller for \{Mn$_4$\} than for \{Co$_4$\}, as can be appreciated also through the small differences in the spatial distribution of the spin density ($\rho_{up} - \rho_{down}$ on the oxygen atoms, Fig.~\ref{spin-density}).
In \{Mn$_4$\} the magnetic moment of bridging oxygens (O$_A$ and O$_B$)  is negligible.
In \{Co$_4$\} the bridging phenoxy oxygens have opposite magnetization: O$_{B2}$ and O$_{B3}$ are magnetized up
while O$_{B2}$ and O$_{B3}$ are magnetized down. 
Out of the eight acetate oxygens O$_A$, only two (O$_{A^+}$) display a small positive magnetic moment while the other six (O$_{A^-}$) have a larger (in modulus) negative magnetic moment (mean value reported in Table \ref{tab-spin}).
The average magnetization of the N atoms is comparable to the average contribution of $O_A$ but with opposite sign. As a consequence, despite the presence of local magnetic moments, the total spin of both  molecules is S$_{TOT}=0$ confirming their antiferromagnetic character.
Nevertheless, the major spread of the magnetic moment observed in \{Co$_4$\} is an indication of a possible large magnetic interaction of this molecule with other systems when the complex is used for functionalization.   

\begin{table}[t]
  \caption{Magnetic moment ($\mu_B$) of the \{Mn$_4$\} and \{Co$_4$\} molecular complexes. Average values are reported for N (variance 0.004 $\mu_B$), and O$_A$ atoms with positive (O$_{A_{+}} = $ O$_{A_{3,7}}$) and negative (O$_{A_{-}} =$ O$_{A_{1,2,4,5,6,8}}$) magnetic moment (variances in \{Mn$_4$\}/\{Co$_4$\} are 0.0/0.001 $\mu_B$ and 0.001/0.002 $\mu_B$, respectively). 
  Note the (anti)ferromagnetic coupling between (non-)equivalent metal ions: 
  M$_1\sim -$M$_3$, M$_2\sim -$M$_4$, M$_1 = $ M$_2$, M$_3 =$ M$_4$.
  }
  \label{tab-spin}
\begin{tabular*}{0.48\textwidth}{@{\extracolsep{\fill}}llllllll}
    \hline
  ($\mu_B$) & $M_{1,2}$  & $M_{3,4}$ & N & O$_{A_{+}}$ &O$_{A_{-}}$ & O$_{B_{2,3}}$ & O$_{B_{1,4}}$\\
    \hline
    \{Mn$_4$\} & 4.82  & -4.89 & 0.01  & 0.01 & -0.01 & 0.00 & 0.00 \\ 
    \{Co$_4$\} &  2.71 & -2.73 & 0.05  & 0.03 & -0.04 & 0.02 & -0.02 \\
    \hline
  \end{tabular*}
\end{table}

\begin{figure}[t]
\centering
  \includegraphics[width=0.48\textwidth]{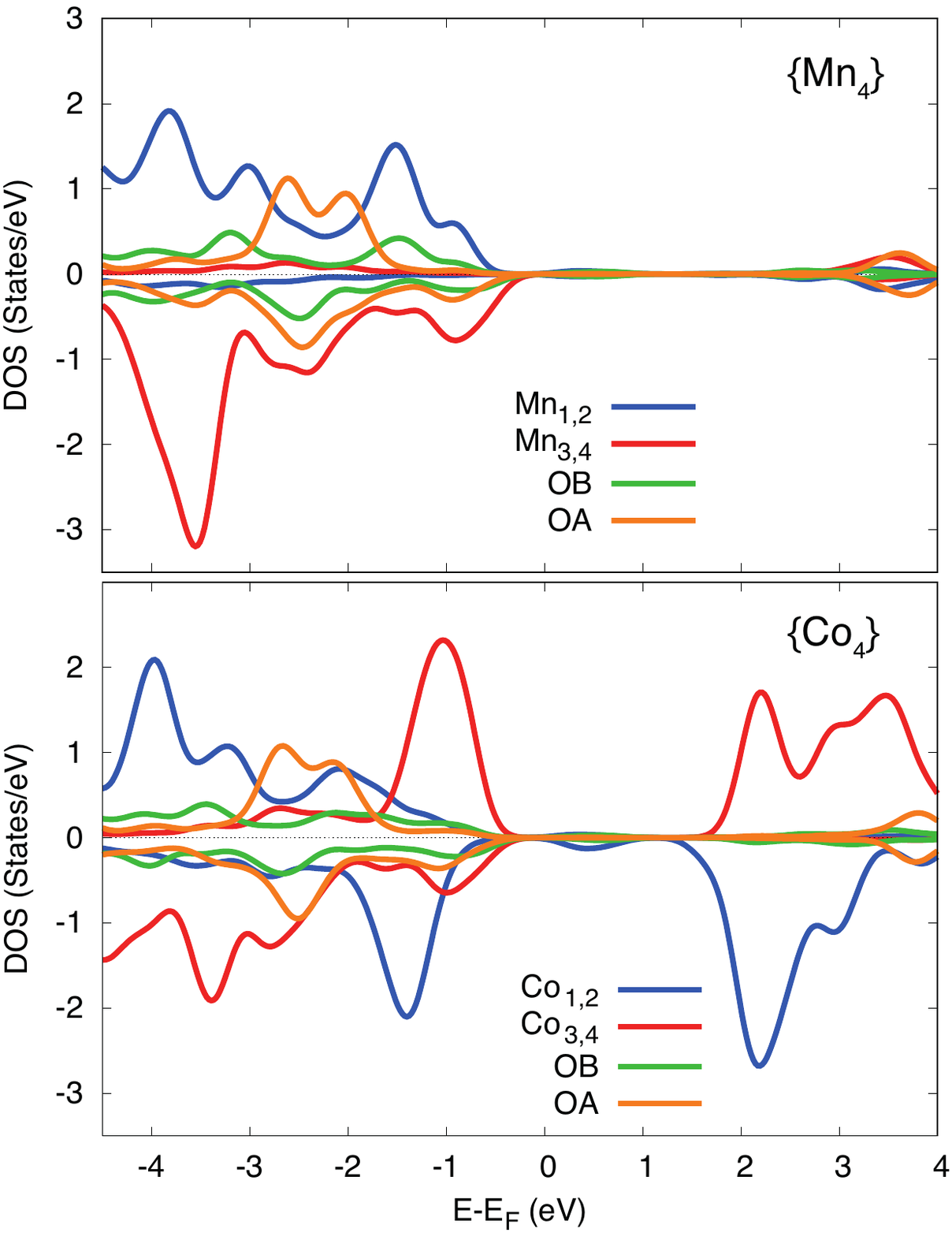}
  \caption{Density of states projected on the magnetic ions and different oxygen atoms (O$_A$ and O$_B$) in the two molecular complexes. The average PDOS per atom type is reported.}
  \label{PDOS}
\end{figure}

In Figure \ref{PDOS} the DOS of the two molecules projected on different atoms of the complex (PDOS), is reported. The PDOS of Mn ions is characterized by a single spin population, due to the almost complete filling of the spin-up $3d$ electrons. In Co, instead, majority and minority spins are present, according to the more-than-half filling of the $3d$ orbitals.
The oxygen atoms display a different PDOS depending on the group they are attached to. In particular the states of the bridging oxygen atoms O$_\textrm{B}$ partially overlap with the states of the magnetic ions, with larger extent for the seven-coordinated ones (M$_{1,2}$) in the majority-spin component and with the six-coordinated ones (M$_{3,4}$) in the minority spin component. This overlap, which appears to be slightly more intense in \{Mn$_4$\}, is responsible for the coupling between magnetic ions via superexchange interaction mechanism.\cite{LaunayVerdaguer}

For both complexes, the O$_\textrm{A}$ atoms are characterized by majority spin states in the [$-3,-1$]~eV energy range and minority states centered around $-2.5$~eV,
with a moderate overlap with the metal atoms in both cases.

We can therefore conclude that they contribute to the magnetic coupling between metal centers with a similar strength.

The hybridization with states of the ligands is also responsible for the charge transfer from the magnetic ions to the nearby atoms. In the molecular complexes, Mn and Co atoms display a number of electrons smaller than the valence of the isolated atom, as reported in Table~\ref{tab-charge} in term of the net atomic charge, i.e. they donate electron charge. This reduction of charge is larger for the two metal ions bound to the pyridine-diimine group (M$_{1,2}$) with respect to the 6-coordinated magnetic atoms in the same molecule (M$_{3,4}$). Moreover, the percentage of lost charge is larger in \{Mn$_4$\} ($\sim 25\%$) than in \{Co$_4$\} ($\sim 22\%$), according to the larger hybridization with the sourrounding coordination groups. 
The oxygen atoms act as electron acceptors in both molecules. The maximum charge transfer is toward the bridging oxygens that acquire 0.60 and 0.68 electrons (mean values) in \{Mn$_4$\} and \{Co$_4$\}, respectively. For O$_A$ atoms the absolute values of the acquired charge ($\sim 0.4 e$) are similar in the two complexes.
N atoms participate to charge transfer towards the ligands by donating electrons, with a slightly larger fraction in \{Mn$_4$\}($\sim 0.29e$) than in \{Co$_4$\} ($\sim 0.21e$). 
On the basis of the results reported by Kampert {\textit et al.} \cite{Kampert09}  
the charge withdrawn from the [M$_4$O$_4$] core (by the ligands) is inversely proportional to the strength of the magnetic interaction within the complex.
For both complexes, the analysis of Mulliken charges predicts that the [M$_4$O$_4$] core acts as a donor, with 2.45 and 3.43 electrons donated in the Mn and Co case, respectively. Therefore, we expect that a stronger antiferromagnetic coupling in \{Mn$_4$\} than in \{Co$_4$\}, supporting the experimental findings.

\begin{table}[t]
  \caption{\ Net atomic charges of the atomic species in the \{Mn$_4$\} and \{Co$_4$\} molecular complexes (in units of electron charge $e$). Positive (negative) values indicate
  donor (acceptor) behavior. Average values are reported for the equivalent centers.
  }
  \label{tab-charge}
  \begin{tabular*}{0.48\textwidth}{@{\extracolsep{\fill}}llllll}
    \hline
   $\Delta q (e)$ & $M_{1,2}$  & $M_{3,4}$ & N & O$_A$ & O$_{B}$ \\
    \hline
    \{Mn$_4$\} & +1.10 & +1.32 & +0.29  & -0.39 &  -0.60 \\ 
    \{Co$_4$\} &  +1.49 & +1.63 & +0.22 & -0.44 &   -0.70 \\
    \hline
  \end{tabular*}
\end{table}

\subsection{Role of Spin Orbit Coupling} \label{SO}
In the previous paragraphs we have analyzed the results obtained with the LDA+U approximation, 
necessary to account for the electronic correlation of localized $3d$ orbitals, but limited in the SIESTA code to a 
collinear-spin description of magnetism.
This approximation is valid for the Mn(II) ions, as the high-spin $d^5$ electronic configuration 
does not have a net orbital momentum. In  Co(II) centers, instead, the $d^7$ configuration leads to an orbital momentum L=3 for an isolated ion. 
While the orbital momentum is quenched in low symmetry environments, including pentagonal bipyramidal, it is not in a perfect octahedron, where L=1. 
As two of the metal centers in \{Co$_4$\} display a pseudo-octahedral geometry, SOC
is expected to have a significant effect on the magnetic properties of this complex.

In order to 
investigate the role of SOC in the complexes, we have performed DFT calculations including SOC 
for both molecules at $U = 0$ using the fully relativistic pseudopotential formalism \cite{Cuadrado12} implemented in SIESTA \cite{Cuadrado21}.
The magnetization direction has been set along the $z$ axis for \{Mn$_4$\}. Indeed we verified that for this molecule the magnetic anisotropy related to spinflip along five independent directions is at most $\sim 100~ \mu$eV per molecule. 
For \{Co$_4$\}, which is expected to display strong spin-orbit effects, we have explored 30 different direction of the magnetization. The easy axis for \{Co$_4$\} is rotated with respect to the $z$ direction with a polar angle 150$^\circ$ and azimuthal angle 45$^\circ$. The maximum magnetic anisotropy for spinflip amounts to $~12$~meV per molecule.
The results reported below are relative to \{Co$_4$\} with the spin along the easy axis.

In Table \ref{tab-SOC1} and \ref{tab-SOC2} the computed S, L, and their sum (J) are reported for both molecular complexes.
In both cases, the atomic spin is slightly reduced with respect to the LDA+U calculation. 
The four metal ions in \{Co$_4$\} display an orbital moment which is smaller than the value expected for the isolated ion, but significantly larger than the manganese analogue. The quenching is stronger for the two ions in the pentagonal bipyramidal coordination, for which $\sqrt{<\rm{L}^2>}=0.11$. For the 6-coordinated pseudo-octahedral Co ions L is not negligible ($\sim 0.4$) and contributes to an overall value of $\sqrt{<\rm{J}^2>} \sim 3$.

On the basis of these results we can infer that  Spin-Orbit coupling plays a crucial role in determining the magnetic properties of \{Co$_4$\}
while it is less relevant in \{Mn$_4$\} and in the latter case it can be disregarded.

\begin{table}[t]
  \caption{Spin (S), orbital moment (L) and their composition (J) for the four magnetic atoms of the \{Mn$_4$\} core. The data are reported in units of $\mu_B$.}
  \label{tab-SOC2}
  \begin{tabular*}{0.48\textwidth}{@{\extracolsep{\fill}}lllll}
    \hline
    & Mn$_1$ & Mn$_2$ & Mn$_3$  & Mn$_4$  \\
    \hline
    $\sqrt{{\rm <S}^2>}$  &  4.33 & 4.37 & 4.34 &  4.34 \\ 
    S$_x$ &  1.43 & -1.35 & 0.17  & 0.37 \\
    S$_y$ & 1.097 & -0.59 & -0.20 &  0.058 \\
    S$_z$ & 3.94  & 4.11 & -4.33 & -4.32  \\ 
    \hline
    $\sqrt{{\rm <L}^2>}$ & 0.052 & 0.057 & 0.058 & 0.058 \\
    L$_x$ & 0.013 & -0.015 & -0.007 & -0.003 \\
    L$_y$ & 0.006 & -0.008 & 0.004 &  0.01 \\
    L$_z$ & 0.05 & -0.054 & -0.057 & -0.057 \\
    \hline
    $\sqrt{{\rm <J}^2>}$ &4.38  & 4.40 & 4.39 & 4.40 \\
  \end{tabular*}
\end{table}

\begin{table}[t]
  \caption{Spin S, orbital moment L and their composition (J) for the four magnetic atoms of the \{Co$_4$\} core.The data are reported in units of $\mu_B$.}
  \label{tab-SOC1}
  \begin{tabular*}{0.48\textwidth}{@{\extracolsep{\fill}}lllll}
    \hline
    & Co$_1$ & Co$_2$ & Co$_3$  & Co$_4$  \\
    \hline
    $\sqrt{{\rm <S}^2>}$  &  2.56 & 2.56 & -2.57 &  -2.58 \\
    S$_x$ &  -0.19 & 1.92 & -1.92  & 0.975 \\
    S$_y$ & 1.355 & 1.6 & -1.62 &  -0.72 \\
    S$_z$ & -2.61  & -0.56 & -0.52 & 2.29  \\ 
    \hline
    $\sqrt{{\rm <L}^2>}$ & 0.15 & 0.14 & 0.46 & 0.35 \\
    L$_x$ & -0.004 & 0.121 & -0.33 & 0.057 \\
    L$_y$ & 0.051 & 0.065 & -0.33 &  0.06 \\
    L$_z$ & -0.14 &-0.033 & -0.033 & 0.33 \\
    \hline
    $\sqrt{{\rm <J}^2>}$ &2.71 & 2.71 & 3.03 & 2.92 \\
  \end{tabular*}
\end{table}

\subsection{Exchange coupling}
\label{sec-J}
In order to ascertain the strength of the (anti)ferromagnetic coupling in the two complexes we computed the exchange coupling parameters ($J_{i}$) from first-principles total energies by considering the five lowest-energy spin configurations of the molecules ($uudd$, $udud$, $uddu$, $uddd$, $uuuu$) as explained in Section \ref{methods}.
The calculated exchange parameters are reported in Table \ref{tab-exchangeSOC}, where positive and negative values correspond to ferromagnetic (FM) or antiferromagnetic (AFM) coupling. 
The parameters have been obtained either in the LDA+U approximation or including SOC in the calculation (with spin along the easy-axis direction).

The LDA+U simulations give $J$ that are of the same order of magnitude of those extracted from experimental susceptibility (see ESI), and are in fair agreement with those reported in Ref. \cite{Kampert09} 
for a three-$J$ model.
Both $J_1$ and $J_2$ are negative, confirming the antiferromagnetic coupling between not equivalent atoms.
One of the other two parameters describing the coupling between equivalent atoms ($J_3$ and $J_4$) is positive (FM), in agreement with the data extracted from the experiments.
The overall exchange interaction, estimated as the average of the $Js$ ($-3.7$ meV for \{Co$_4$\} and $-1.4$ for \{Mn$_4$\}), is AFM for both  molecular complexes.

The strongest $J_n$ ($J_1$ in  \{Co$_4$\}, $J_2$ in  \{Mn$_4$\}) corresponds to the interaction between the two pairs of not-equivalent ions and it is related to the energy difference between the AFM ground state and the high spin FM state (S=12 for \{Co$_4$\} and S=20 for \{Mn$_4$\}) which is larger for \{Co$_4$\}.
The FM interaction between equivalent ions (intra-pair)
is, instead, smaller for \{Co$_4$\}.
The latter governs the transition to low-spin FM states (for example $uddd$) which influences the behavior of the magnetization at low fields and low temperatures, hence explaining the observed 
switching with temperature of the magnetization curves M(B) of the two molecules (Fig.~\ref{SQUID}).
To further explore this behavior, we have exploited a model Heisenberg Hamiltonian with $J$ parameters and g-factor fitted from the experimental low-field susceptibility (ESI) and used them to calculate M(B) at two different temperatures. We find that the observed (Fig.~\ref{SQUID}) switch of M(B) with temperature is an effect of the more marked AFM character of \{Mn$_4$\} 
giving rise to a positive curvature of M(B) at low field/low T.  
At high field/high T, instead, the most relevant factor is the larger saturation value of the magnetization in \{Mn$_4$\} compared to \{Co$_4$\}.
By increasing the range of the magnetic field beyond the experimental one, a crossing of the two theoretical curves is observed due to the combination of these two aspects (ESI).

The inclusion of SOC in the calculation leads to
$J$s  with a sign that reflects the $uudd$ magnetic order of the ground state, i.e. AFM (FM) coupling between M1/M4 and M2/M3 non-equivalent (M1/M2 and M3/M4 equivalent) ions. 
Nevertheless, the exchange parameters obtained with SOC are too large compared to those extracted from the experiments, suggesting that the electronic correlation can not be neglected for a reliable estimate of the strength of the magnetic interactions. 

For a deeper insight of the exchange interaction, we have computed the exchange parameters also with the Liechtenstein-Katsnelson-Antropov-Gubanov (LKAG) formula\cite{LKAG1987} implemented in the TB2J package\cite{he2021tb2j}, which treats the local spin rotation of the numerical atomic orbitals for the magnetic atoms as a perturbation. 
The $Js$ evaluated with this approach present the same overall trend as those computed from total energies (Table~\ref{tab-exchangeSOC}), for both LDA+U and SOC case, and are reported in the ESI.

The different predictions of the various approximations (LDAU or SOC) are a consequence of the complexity of the magnetic potential energy landscape of these molecular complexes. A small perturbation (geometry, electron correlation, spin alignment) can drive the results out of equilibrium and towards a different local minimum. Despite these difficulties, all the computed $J$ parameters 
predict the experimentally observed $uudd$ ground state, regardless of the approach (total energies or perturbative) and  inclusion of correlations or SOC. 
This has been verified by feeding the computed $J$ in the Heisenberg Hamiltonian and computing the various spin configurations (ESI).

Finally, we used all the computed $J$ parameters as input for spin dynamics simulations\cite{landau1992theory,gilbert2004phenomenological,gonze2020abinit} to predict
the magnetic susceptibility, finding a good agreement with experiment in the LDA+U case (ESI). The results obtained without U correction are, instead, in striking contradiction with experiments, confirming the importance of taking into account electron correlation in the transitional metal sites.

\begin{table}[t]
  \caption{Exchange coupling parameters ($J_i$, meV) of \{Mn$_4$\} and \{Co$_4$\} molecular complexes extracted from DFT calculations with either LDA+U or SOC included. S is normalized to 1.}
  \label{tab-exchangeSOC}
  \begin{tabular*}{0.48\textwidth}{@{\extracolsep{\fill}}lllll}
    \hline
    $LDA+U$ (meV) & $J_1$ & $J_2$   & $J_3$  &  $J_4$\\
    \hline
  
    \{Mn$_4$\} & -0.2  & -0.9 & -0.2 & 0.9  \\ 
    \{Co$_4$\} & -1.8 & -0.19 & 0.8  & -0.6  \\
    \hline
      \hline
    $SOC$ (meV) & $J_1$ & $J_2$   & $J_3$  &  $J_4$\\
    \hline
    \{Mn$_4$\} & -5.0 & -4.2 & 9.3 & 31.8 \\ 
    \{Co$_4$\} & -7.35 & -1.42 & 2.58  & 5.52  \\
    \hline
  \end{tabular*}
\end{table}

\section{Conclusions}
Through a joint experimental and theoretical analysis we have characterized the properties of two coordination complexes, \{Mn$_4$\} and \{Co$_4$\}, that display the same chemical structure but different inner magnetic core formed by Mn and Co atoms, respectively.
The theoretical analysis was performed under different approximations (LDAU, SOC) and methods (first-principles,  model Hamiltonians, perturbation theory).

The experimental data and the theoretical calculations show that by changing the magnetic core it is possible to tune the strength of the magnetic interaction inside the molecules and thus the robustness of the AFM configuration in an external magnetic field. 
Moreover, the different magnetic properties of the two chemical species lead to a different spatial extension of the magnetic moment and electronic charge density on the ligands, which influences the interaction with foreign systems and affects the efficiency of the two compounds when employed for magnetic functionalization.

We find and explain an unusual switch with temperature of the dependence of the magnetic moment from the applied magnetic field M(B)
for the two molecules. We relate it to two competing effects:
the stronger AFM coupling in \{Mn$_4$\}  
and the large value of magnetization in \{Mn$_4$\}
which dominate at low and high temperature, respectively.

The calculations clarify the role of spin-orbit effects: negligible in \{Mn$_4$\} and relevant in \{Co$_4$\}, showing that SOC has to be considered for a reliable theoretical description of the magnetic moments of the latter.
In perspective of future exploitation of these compounds in spintronics the SOC effects found in \{Co$_4$\} should be taken into account as possible source of spin decoherence.

Our study of the exchange coupling parameters and spin dynamics demonstrate that it is necessary to explicitly include electron correlations (for instance, via a Hubbard U parameter) to properly recover these properties. 
The complete description of the molecular complexes can only be performed in a framework in which electronic correlation and SOC are treated on the same footing.

\section*{Conflicts of interest}
There are no conflicts to declare.

\section*{Acknowledgements}
The authors thank Natalya Izarova for acquisition of the crystallographic data, Brigitte Jansen for acquisition of the TGA data and Christina Houben for acquisition of some of the SQUID data.
The Authors acknowledge financial support of the NFFA infrastructure under Project ID-753. Computational resources were provided by the Red Espanola de Supercomputation through a the projects FI-2019-2-0038 and FI-2020-1-0022 on Marenostrum High Performance cluster. We acknowledge PRACE for awarding us access to MareNostrum4 at Barcelona Supercomputing Center (BSC), Spain (OptoSpin project id. 2020225411).
ZZ acknowledges financial support by the Ramon y Cajal program RYC-2016-19344 (MINECO/AEI/FSE, UE) and the Netherlands Sector Plan program 2019-2023. PO, HX and ZZ thank the support by the EU H2020-NMBP-TO-IND-2018 project ”INTERSECT” (Grant No. 814487), the EC H2020-INFRAEDI-2018-2020 MaX ”Materials Design at the Exascale” CoE (grant No. 824143), Spanish AEI Grant Fis2015-64886-C5-38, Severo Ochoa (SEV-2017-0706) and Generalitat de Catalunya (CERCA program and Grant 201756R1506).


\balance


\bibliography{main} 
\bibliographystyle{rsc} 

\end{document}


\maketitle

\noindent
\footnotesize{\textit{$^{a}$}~Dipartimento di Fisica "Aldo Pontremoli", Universit\'a degli Studi di Milano, Via Celoria 16, Milan, Italy, simona.achilli@unimi.it}

\noindent
\footnotesize{\textit{$^{b}$}~European Theoretical Spectroscopy Facilities.}

\noindent
\footnotesize{\textit{$^{c}$}~Department of Chemistry, The George Washington University, Washington DC 20052, USA.}

\noindent
\footnotesize{\textit{$^{d}$}~Catalan Institute of Nanoscience and Nanotechnology (ICN2), CSIC and BIST, Campus UAB. }

\noindent
\footnotesize{\textit{$^{e}$}~Department of Physics, Universität Osnabrück, 49076 Osnabrück, Germany.}

\noindent
\footnotesize{\textit{$^{f}$}~Chemistry Department, Debye Institute for Nanomaterials Science, Condensed Matter and Interfaces, Utrecht University, PO Box 80 000, 3508 TA Utrecht, The Netherlands.}

\vspace{1cm}

\section{Synthesis and characterization}

\subsection{Materials and synthesis}
2-aminophenol ($>98\%$, TCI) was recrystallised from boiling water, washed with ice-cold water and diethylether and stored under inert atmosphere. 2,6-diacetylpyridine ($>98\%$, TCI), Co(OAc)$_2\cdot$4 H$_2$O ($99.999\%$ trace metal basis, Aldrich) and Zn(OAc)$_2\cdot$2 H$_2$O (99.999\% trace metal basis, Aldrich) were used as received. Solvents were, when indicated, deaerated by three freeze-pump-thaw cycles using argon as inert gas. PE spatulas and PTFE cannulas were employed to avoid magnetic contamination of the samples.

\subsubsection*{[Mn$_4$OAc$_4$L$_2$] (H$_2$L=2,6-bis-(1-(2-hydroxy\-phenyl)imino\-ethyl)pyridine) (\{Mn$_4$\})} 
The complex was synthesized according to the literature.\cite{Kampert09} 

\subsubsection*{[Co$_4$L$_2$(OAc)$_4$] (H$_2$L = 2,6-bis-(1-(2-hydroxy\-phenyl)imino\-ethyl) pyridine) (\{Co$_4$\})} Synthesis of the complex follows the preparation of the manganese analogue,\cite{Kampert09} with some modifications. In particular, conducting the reaction under protective atmosphere is necessary to avoid the formation of an unidentified brown precipitate by the reaction of 2-aminophenol, cobalt acetate and dioxygen. 2,6‑diacetylpyridine (439 mg, 2.5 mmol, 1 eq.), 2-aminophenol (583 mg, 5.3 mmol, 2.2 eq.) and Co(OAc)2·4 H$_2$O (1.33 g, 5.3 mmol, 2.2 eq.) were introduced in a Schlenck flask under argon. Deaerated methanol (10 mL) was canulated into the flask and the resulting red solution was refluxed under argon for two hours. After the solution was returned to room temperature, deaerated diethylether (100~mL) was canulated in and allowed to mix with the solution, yielding an orange microcrystalline precipitate. The solid was filtered and washed with diethylether (3$\times$30 mL) to yield Co$_4$L$_2$(OAc)$_4\cdot$4 CH$_3$CN (1.0-1.2 g, 0.8-0.9 mmol, 60-70 \%) as a dark orange powder. The compound was recrystallised by dissolving the solid (130 mg, 0.1 mmol) in acetonitrile (25 mL), sonicating and filtering solution and setting it up for gas phase diffusion of diethylether. Dark red crystals of Co$_4$L$_2$(OAc)$_4\cdot$4 CH$_3$CN (65-100 mg, 50-75 \% recrystallisation yield) were collected by filtration after 3-4 days and used for all further characterization and reactions.

UV (CH$_3$CN) $\lambda_{max}$, nm ($\log\epsilon$): 260sh (4.50), 321 (4.27), 398 (4.14).

CV $E_{1/2}$, V vs. Fc$^+$/Fc: 0.30, 0.80.

TGA (N$_2$, 10 K.min$^{-1}$) -12.3 \% (50-150 K, -4 CH$_3$CN (th. -12.4 \%)). -30.5 \% (350-400 K), -9.4 \% (400-520 K).

Single crystal diffraction quality material can be obtained by gas phase diffusion of \emph{n}-pentane into a saturated ethanol solution. The compound crystallizes in the P4/n space group with ethanol as a solvate. Due to twinning of the crystals and disorder of the solvate molecules we were only able to refined to a $wR^2$ of 36 \%. The asymmetric unit unequivocally contains a Co$_4$L$_2$(OAc)$_4$ molecule and an ethanol molecule hydrogen bound to one of the acetate oxygens, as well as a second ethanol molecule which refines satisfactorily with a two-parts disorder of the hydroxyl group with approximatively equal occupancies. The SQUEEZE routine was used to account for the residual electronic density localised in the large voids of the unit cell, which corresponds to 8 ethanol molecules per unit cell, yielding a final $wR^2$ of 36 \% ($R_1=18 \%$) with a formula Co$_4$L$_2$(OAc)$_4\cdot$3 EtOH. Crystal structure data is available from Cambridge Structural Database, under CCDC number 1855019.

\subsubsection*{Zn$_4$L$_2$(OAc)$_4$ (H$_2$L = 2,6-bis-(1-(2-hydroxy\-phenyl)imino\-ethyl) pyridine) (\{Zn$_4$\})} Synthesis of the complex follows the preparation of the manganese analogue,\cite{Kampert09} with some modifications. Specifically, 2,6-diacetylpyridine (186 mg, 1.0 mmol, 1 eq.), 2-aminophenol (250 mg, 2.3 mmol, 2.3 eq.) and Zn(OAc)$_2\cdot$2 H$_2$O (501 g, 2.3 mmol, 2.3 eq.) were introduced in a Schlenck flask under argon. Deaerated methanol (10 mL) was canulated into the flask and the resulting yellow solution was refluxed under argon for two hours. Note that this step can be conducted in air, but in that case a small amount of strongly red-colored 2-amino-3H-phenoxazin-3-one is formed by the oxidative dimerisation of 2‑aminophenol and contaminates the product obtained before recrystallisation. After the solution was returned to room temperature, a microcrystalline yellow solid was precipitated by the addition of diethylether (200~mL). The solid (590-640~mg) was filtered, washed with diethylether (4$\times$25~mL), and redissolved in chloroform (100 mL). Pentane was allowed to diffuse in the solution from the gas phase (to accelerate the process, the solution was divided in three fractions). Yellow single crystal diffraction quality Zn$_4$L$_2$(OAc)$_4\cdot$4.5CHCl$_3$ (350-400~mg, 40-45~\% yield) was collected by filtration after 3-4 days and used for all further characterization and reactions.

The compound crystallised in the P2$_1$/n space group, with 4.5 CHCl$_3$ molecules and a Zn$_4$L$_2$(OAc)$_4$ complex in the asymmetric unit. One of the chloroform molecules is located close to the inversion center at (0, 0.5, 0) and has an occupancy of 0.5. Another could be refined as an approximately 1:1 disorder between two close positions. Residual electronic density lies mostly in close proximity with the chlorine atoms of the other chloroform molecules, but attempts at refining similar disorder yielded only small occupancy factors and no significant improvement of the $R$ factors, and were abandoned, leaving a final $wR^2$ factor of 12 \% ($R_1=4.3\%$). Crystal structure data is available from Cambridge Structural Database, under CCDC number 1855020.

UV (CH$_3$CN) $\lambda_{max}$, nm ($\log\epsilon$): 265sh (4.40), 293 (4.23), 346 (4.17), 407 (4.23).

SQUID: $\chi_{dia}$, m$^3$.mol$^{-1}$ (emu.mol$^{-1}$): -5.3$\times$10$^{-9}$ (4.2$\times$10$^{-4}$).

\newpage
\section{Fit of SQUID data}
\begin{figure}[t]
\centering
  \includegraphics[width=0.7\textwidth]{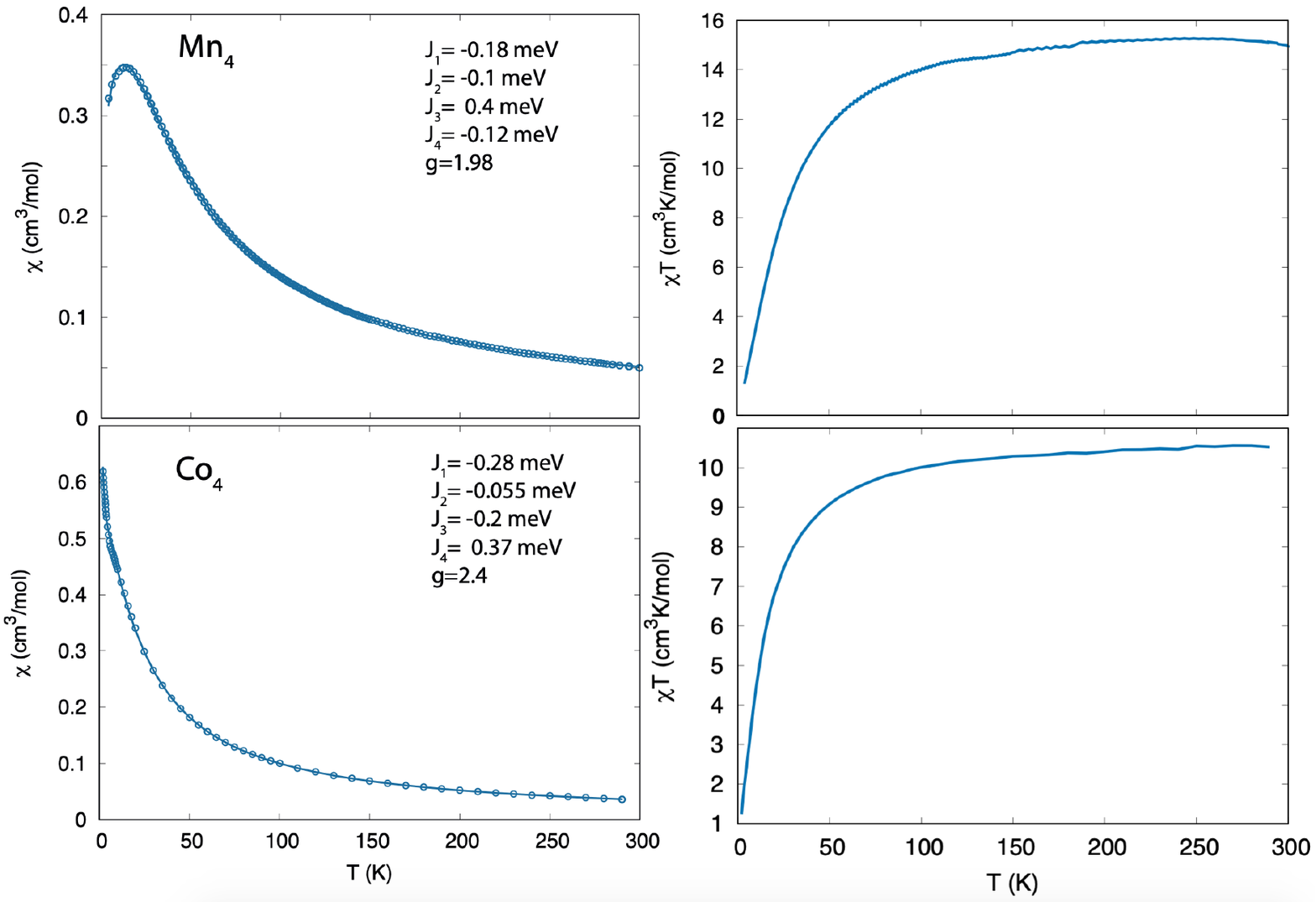}
  \caption{Left panels: Experimental (dots) magnetic susceptibility ($\chi$ cm$^3$/mol) of \{Mn$_4$\} and \{Co$_4$\} with the relative fit (line). The $J$ derived from fitting are reported as inset together with the value of $g$. Right panels: experimental susceptibility times temperature ($\chi \times T$, cm$^3$K/mol) of \{Mn$_4$\} and \{Co$_4$\} as a function of the temperature.  
   }
  \label{Susc}
\end{figure}

The experimental magnetic susceptibility
was fitted exploiting an Heisenberg model Hamiltonian in which the exchange coupling parameters between pairs of ions are left as free parameters.
The fit functions are reported in Figure \ref{Susc} as solid lines, together with the experimental data (points) and the $J_n$ extracted from the fit.

The $J_n$ obtained are in fair agreement with those extracted from the LDA+U calculation.
The fitted g-factor is $\sim 2$ in \{Mn$_4$\} confirming the validity of the spin-only approximation, while it deviates from 2 in \{Co$_4$\} reflecting the role of SOC.

Through the model Hamiltonian with parameters obtained from the fit of the experimental data the magnetization was calculated at different temperatures and fields.
For large fields or large temperatures the magnetization of \{Mn$_4$\} is larger than that of \{Co$_4$\}, as observed in the experiments. The saturation value ($gS\mu_B$) of \{Mn$_4$\} ($20 \mu_B$) is indeed larger than for \{Co$_4$\} ($12 \mu_B$ in a spin-only model). At low temperature, fields of 20 T for \{Co$_4$\} and 25 T for of \{Mn$_4$\} are needed for the magnetization to reach saturation. These values  are significantly large compared to the fields  accessible in our experiments (Fig. 3 in the main text). 

In the low field$-$low-temperature limit the model shows an inversion of the magnetization curves leading to larger magnetization of \{Co$_4$\} with respect to \{Mn$_4$\}. Moreover the two curves display opposite curvature, in agreement with SQUID data at low temperature. In particular, the magnetization of \{Mn$_4$\} shows positive curvature, as expected for an antiferromagnet, while for \{Co$_4$\} an almost linear dependence on the field can be observed.


The slope of the magnetization is an indication of the coupling between the magnetic centers: the stronger the antiferromagnetic coupling, the smaller the slope. Moreover, the smaller the moment per atom, the stronger the linear dependence of M on the field.
Both these aspects explain the fast rise observed for the magnetization of \{Co$_4$\} in the experimental data, as due to the smaller value of S (or J) (see table 4 and 5 in the main text) and of the AFM coupling, with respect to \{Mn$_4$\}. On the other hand, the saturation value for the magnetization of \{Co$_4$\} is smaller than for \{Mn$_4$\}, thus the two magnetization curves show a crossing for a certain value of the field which is not reached in the experiments.
At high temperature the magnetization curves flatten,
resulting in a slower growth as a function of the field with respect to the low temperature limit. The linear dependence on the field $M=Ng^2{\rm S(S+1)}\mu_B^2B$ in this case is governed by the value of S and it is smaller for \{Co$_4$\} than for \{Mn$_4$\}, in agreement with what observed in the experiments at high temperature.

\begin{figure}[h]
\centering
  \includegraphics[width=0.48\textwidth]{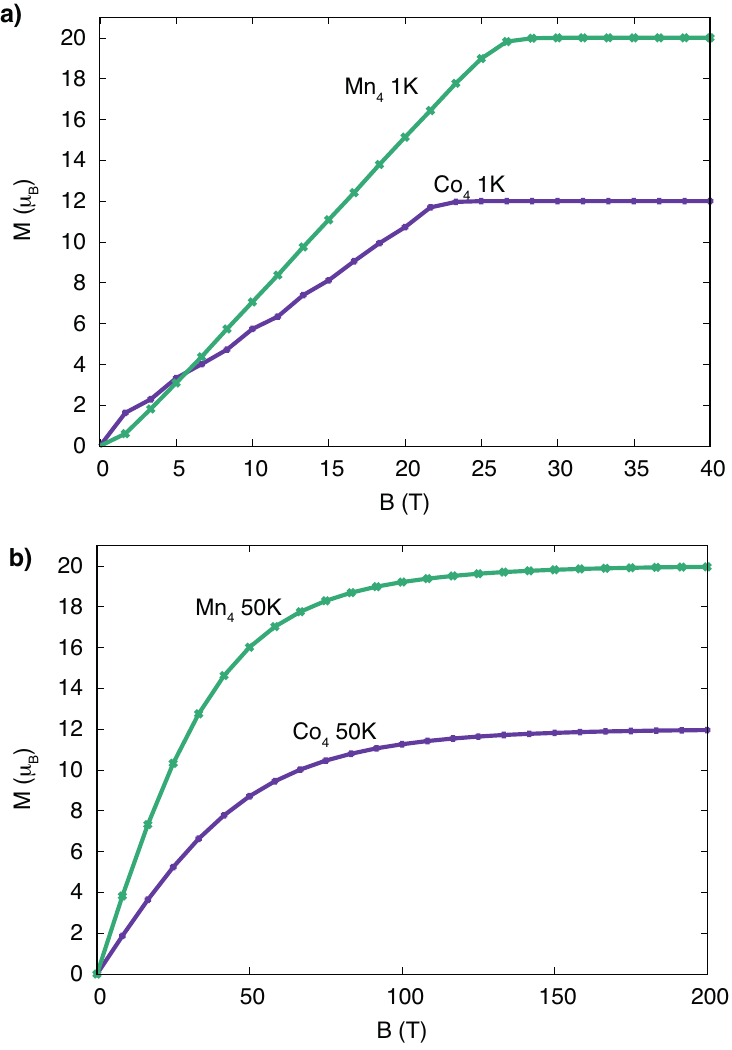}
  \caption{Magnetization versus field dependence for \{Mn$_4$\} and \{Co$_4$\} at 1 K (a) and 50 K (b).}
  \label{mag-fit}
\end{figure}

\newpage

\section{Exchange parameters from different calculations}
The exchange parameters where extracted by fitting a system of equation for the five lowest energy spin configurations of the molecule, as reported in the main text.
For a comparison with the available experimental data and previous theoretical calculations on \{Mn$_4$\} we consider also a set of only three $J$ parameters, despite we retain this choice less accurate. Indeed the assumption $J_3 = J_{(1,2)} = J_{(3,4)}$ is not correct, as the M$_1$/M$_2$ and M$_3$/M$_3$ pairs are not equivalent in terms of chemical coordination and ligands groups. 
The obtained values are reported in Tables \ref{tab-3exchange}
For a system of three exchange parameters, we recover the experimental results of Kampert {\it et al.} ~\cite{Kampert09}.

In addition, we also computed the $J$ values with with the Liechtenstein-Katsnelson-Antropov-Gubanov (LKAG) formula\cite{LKAG1987} implemented in the TB2J package. \cite{he2021tb2j}

\begin{table}[h]
\small
  \caption{Computed exchange parameters J (meV and Kelvin) for the \{Mn$_4$\} molecular complex, using LDA+U and a three-J Heisenberg model. The experimental values from Ref.~\cite{Kampert09} are reported for comparison}
\label{tab-3exchange}
  \begin{tabular*}{0.8\textwidth}{@{\extracolsep{\fill}}llll}
    \hline
    $J$                & $J_1$ (1-4,2-3) & $J_2$ (1-3,2-4)  & $J_3$ (1-2, 3-4) \\
   \hline
   \{Mn$_4$\} (meV)             & -0.7  & -0.4  & -0.2  \\ 
   \{Mn$_4$\} (Kelvin)          & -8.1  & -4.1  & -2.0   \\ 
   Exp~\cite{Kampert09}(Kelvin) & -2.2  & -1.1  & -0.1   \\
    \hline
  \end{tabular*}
\end{table}

\begin{table}[h]
\small
  \caption{J of the \{Mn$_4$\} and \{Co$_4$\} molecular complexes obtained with TB2J in LDA+U and with SOC included. The values are reported in meV.}
\label{TB2J1}
  \begin{tabular*}{0.8\textwidth}{@{\extracolsep{\fill}}lllll}
    \hline
    LDA+U  & $J_1$ (1-4,2-3) & $J_2$ (1-3,2-4)  & $J_3$ (1-2) &  $J_4$ (3-4)\\
   \hline
   \{Mn$_4$\} & -1.21/-1.23  & -1.77/-1.84 & -0.8 & 0.1  \\ 
    \{Co$_4$\} & -3.47/-3.29 & -1.70/-1.82 & 0.03  & -2.25  \\
    \hline
    \hline
    SOC & $J_1$ (1-4,2-3) & $J_2$ (1-3,2-4)  & $J_3$ (1-2) &  $J_4$ (3-4)\\
   \hline
   \{Mn$_4$\} & -34.5/1.3  & -55.7/-29.2 & 12.94 & 3.29  \\ 
    \{Co$_4$\} & -3.24/-5.09 & -11.84/-10.68 & 1.18  & 18.35  \\
    \hline
  \end{tabular*}
\end{table}

TB2J evaluates the J for each couple of atoms. So there are six J parameters. 
The non-equivalent $J$ for pairs M$_1$/M$_3$ and M$_2$/M$_4$ is related to the not perfect symmetry of the electronic structure associated with these pairs of atoms, probably due to some orbital polarization in the Mn/Co.

When the Heisenberg equation is solved for four exchange parameters, The TB2J results for the LDA+U case give the same trend obtained using SIESTA total energies:
the ferromagnetic coupling between atoms M1 and M2 is not reproduced ($J_3 <0$)
and the exchange coupling parameters are larger for \{Co$_4$\} than for \{Mn$_4$\}
With SOC included, the exchange parameters obtained with TB2J are in fair agreement with those reported in the main text supporting the robustness of the results. The comparison between the two molecules confirms indeed the larger strength of the magnetic coupling in \{Mn$_4$\} with respect to \{Co$_4$\} but the absolute value of the $J$ is too large.

All the methods agree in the qualitative description of the magnetic coupling within the two molecules and in these terms, i.e. only qualitatively, these parameters should be taken into account.
\newpage

\section{Spin dynamics calculation}
Using the Heisenberg model constructed from DFT results, we perform Ginzberg-Landau-Lifshitz (GLL) atomic spin dynamics \cite{landau1992theory,gilbert2004phenomenological} (ASD) to compute the magnetic susceptibility with the MULTIBINIT code\cite{gonze2020abinit}. 
The isothermal magnetic susceptibility is calculated with the fluctuation-dissipation theorem\cite{landau2021guide}:

\begin{equation}
 \chi = \frac{1}{k_B T} (\left < m^2 \right>-\left< m\right>^2) \,,    
\end{equation}
where $\left< \right>$ means the average over the ensembles, $m$ is the magnetization of the system.  

\begin{figure}[h]
    \centering
    \includegraphics[width=0.8\textwidth]{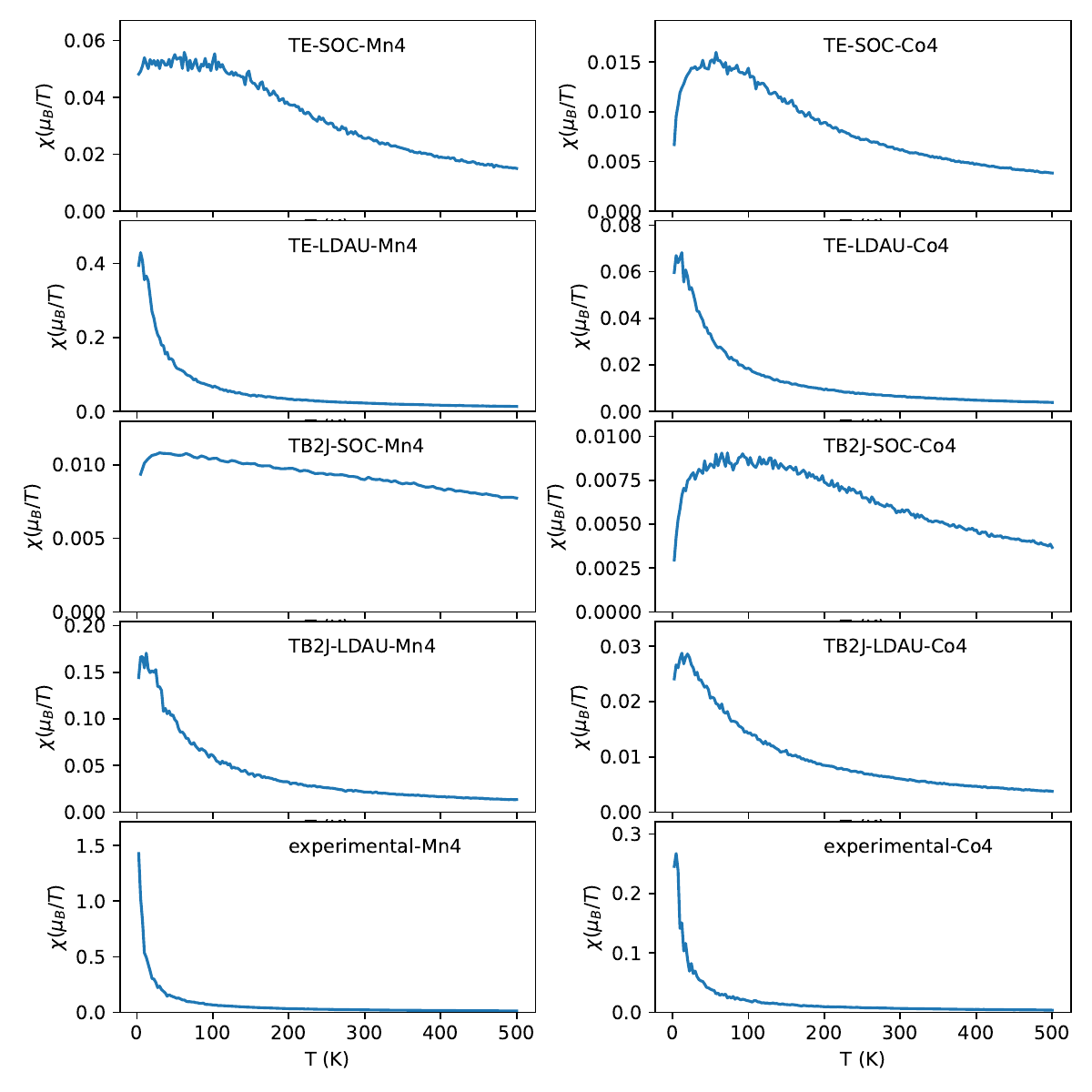}
    \caption{The susceptibilities computed with spin dynamics using various set of parameters. TE-SOC: fit from DFT (with SOC) energies. TE-LDAU: fit from DFT (with +U correction). TB2J-SOC: computed with TB2J from DFT with SOC. TB2J-LDAU: computed from TB2J with DFT+U. Experimental: fit from experimental data.
    The left and the right panels are from Mn4 and Co4, respectively.
    }
    \label{fig:chi_computed}
\end{figure}

Four sets of parameters from DFT results with either SOC or Hubbard U correction, computed by total energy (TE) method or TB2J, were used as input. Another set of exchange parameters by fitting to experimental data is also taken as comparison. The results are shown in Fig.~\ref{fig:chi_computed}.  

The "$\lambda$"-shape of the susceptibility curves are similar to extended structures, where the susceptibility diverges at the critical temperature, which is ill-defined in molecules. Nevertheless the temperature of the peak (definded as $T_P$), the decay of the susceptibility above the $T_P$, and the value of the susceptibilities, all shows that the exchange parameters from the DFT results with the "+U" correction agrees better with experiments, in which the exchange values are smaller in general.

\newpage

\section{Relaxed structures}
The atomic coordinates (in \AA) obtained through structural relaxation performed in DFT-LDA and SOC included are reported below for 
\{Co$_4$\} and \{Mn$_4$\}.

\{Co$_4$\}:\\
Co     11.784660   24.970504    9.657395	\\
Co     10.497270   22.083542   10.349018	\\
Co     13.826910   22.493219    9.595356	\\
Co     11.640250   22.424358    7.355454	\\
O      12.276659   23.190641   10.768667	\\
O      13.053111   23.738692    8.233238	\\
O      12.150853   21.350956    8.965670	\\
O      10.536075   23.509294    8.730816	\\
N      11.150917   25.250576   11.697139	\\
N      10.234291   26.256309    9.631992	\\
N      11.898964   25.955696    7.819901	\\
C      12.573409   23.459183   12.047217	\\
C      13.501743   22.727320   12.800231	\\
H      14.089320   21.934242   12.292846	\\
C      13.683728   23.040467   14.147378	\\
H      14.424537   22.471937   14.743995	\\
C      12.967844   24.084895   14.745447	\\
H      13.125847   24.329670   15.812985	\\
C      12.094520   24.862400   13.986050	\\
H      11.619728   25.748710   14.442161	\\
C      11.894341   24.565336   12.624100	\\
C      10.126773   26.025442   11.948916	\\
C       9.578037   26.345785   13.287772	\\
H      10.175627   27.151091   13.781884	\\
H       8.530036   26.715922   13.221056	\\
H       9.604646   25.444738   13.951800	\\
C       9.538585   26.565027   10.740648	\\
C       8.341386   27.295098   10.668957	\\
H       7.820799   27.608225   11.593808	\\
C       7.833021   27.630308    9.415343	\\
H       6.862677   28.154484    9.327204	\\
C       8.583412   27.341326    8.274367	\\
H       8.239272   27.639455    7.265994	\\
C       9.816490   26.700082    8.429924	\\
C      10.790725   26.483101    7.373647	\\
C      10.499096   26.837914    5.965518	\\
H      11.118355   26.205940    5.283527	\\
H       9.420716   26.654751    5.742013	\\
H      10.706556   27.914849    5.751836	\\
C      13.024326   25.741449    7.058258	\\
C      13.537168   26.594422    6.071137	\\
H      13.031246   27.555676    5.858676	\\
C      14.722039   26.250453    5.422129	\\
H      15.143996   26.930256    4.657679	\\
C      15.384636   25.058139    5.746152	\\
H      16.319763   24.794952    5.214912	\\
C      14.880653   24.193928    6.719788	\\
H      15.372314   23.234492    6.981271	\\
C      13.696788   24.537235    7.377188	\\
N      11.408346   20.414618   11.203267	\\
N      10.045893   22.163800   12.313801	\\
N       8.804952   23.397905   10.567973	\\
C      12.292475   20.027330    9.077568	\\
C      12.842500   19.207261    8.088787	\\
H      13.105214   19.659644    7.110727	\\
C      13.063406   17.859247    8.376878	\\
H      13.502886   17.200660    7.602911	\\
C      12.747203   17.325275    9.633942	\\
H      12.915108   16.251224    9.840454	\\
C      12.205322   18.140545   10.625957	\\
H      11.907104   17.720989   11.605658	\\
C      11.969344   19.493931   10.348564	\\
C      11.650265   20.474168   12.485497	\\
C      12.636444   19.653283   13.225367	\\
H      13.450079   19.328860   12.531885	\\
H      13.088399   20.253265   14.051051	\\
H      12.169384   18.748781   13.685771	\\
C      10.832454   21.468847   13.158681	\\
C      10.796516   21.686320   14.539889	\\
H      11.495754   21.142632   15.202811	\\
C       9.837013   22.560233   15.053198	\\
H       9.769746   22.736793   16.143428	\\
C       8.934736   23.169554   14.182981	\\
H       8.097913   23.779891   14.572506	\\
C       9.094103   22.979726   12.800664	\\
C       8.331740   23.643350   11.763572	\\
C       7.163003   24.495043   12.088568	\\
H       6.207428   23.931855   11.948397	\\
H       7.186354   24.831516   13.149891	\\
H       7.130370   25.394655   11.423137	\\
C       8.324216   23.904560    9.387352	\\
C       7.023869   24.354608    9.088815	\\
H       6.233333   24.332011    9.859428	\\
C       6.706528   24.748870    7.789651	\\
H       5.679835   25.092977    7.561875	\\
C       7.661627   24.652798    6.770087	\\
H       7.379020   24.906744    5.729114	\\
C       8.956450   24.210011    7.041607	\\
H       9.719280   24.084856    6.245056	\\
C       9.303374   23.877674    8.358257	\\
O      13.550916   25.687501   10.333671	\\
O      14.950837   23.896252   10.435570	\\
C      14.509448   25.023697   10.844952	\\
C      15.140291   25.589838   12.078773	\\
H      16.108362   25.094580   12.300594	\\
H      15.260569   26.691401   11.992143	\\
H      14.438239   25.391374   12.931510	\\
O       9.264080   21.016327    9.158996	\\
O       9.981759   21.377015    7.029957	\\
C       9.077203   21.170511    7.908653	\\
C       7.656352   21.159999    7.435931	\\
H       7.600703   21.151344    6.327734	\\
H       7.105154   20.298050    7.871949	\\
H       7.167930   22.096305    7.813696	\\
O      14.512862   20.887840   10.767680	\\
O      15.138232   21.186628    8.688865	\\
C      15.122061   20.452460    9.730707	\\
C      15.736798   19.099933    9.768206	\\
H      16.570752   19.090335   10.507721	\\
H      14.961592   18.371446   10.113626	\\
H      16.110875   18.800866    8.767205	\\
O      12.909671   21.999680    5.790142	\\
O      11.488189   23.663316    5.667504	\\
C      12.484160   23.030674    5.173137	\\
C      13.128020   23.549145    3.938469	\\
H      14.056502   22.988709    3.703249	\\
H      12.409620   23.476724    3.089349	\\
H      13.369196   24.629938    4.096111	\\

\{Mn$_4$\}\\
Mn     11.810113   24.985996    9.743006	\\
Mn     10.424819   22.105112   10.320622	\\
Mn     13.969259   22.519930    9.574551	\\
Mn     11.607671   22.312376    7.271543	\\
O      12.349963   23.147982   10.831958	\\
O      13.140464   23.571435    8.098263	\\
O      12.345050   21.326161    8.840598	\\
O      10.496148   23.528278    8.643963	\\
N      11.103152   25.197501   11.777895	\\
N      10.216281   26.274131    9.676366	\\
N      11.890695   25.881198    7.790930	\\
C      12.598813   23.438037   12.113694	\\
C      13.538059   22.730289   12.880143	\\
H      14.135978   21.938064   12.381465	\\
C      13.716597   23.053512   14.225799	\\
H      14.466366   22.503544   14.828710	\\
C      12.981132   24.093676   14.810091	\\
H      13.138994   24.358651   15.873515	\\
C      12.083104   24.839617   14.047557	\\
H      11.583453   25.716735   14.497662	\\
C      11.870084   24.527046   12.689523	\\
C      10.080681   25.994708   12.005918	\\
C       9.546066   26.368410   13.338673	\\
H      10.158648   27.184771   13.798833	\\
H       8.501386   26.750637   13.269521	\\
H       9.564052   25.494243   14.036190	\\
C       9.519736   26.565525   10.790827	\\
C       8.322393   27.297096   10.700291	\\
H       7.771295   27.596526   11.612739	\\
C       7.835847   27.640072    9.438633	\\
H       6.864361   28.162054    9.339557	\\
C       8.588967   27.345740    8.297761	\\
H       8.238350   27.636652    7.288881	\\
C       9.816809   26.695904    8.461386	\\
C      10.779240   26.441424    7.388195	\\
C      10.469343   26.827821    5.989159	\\
H      11.084954   26.222238    5.280743	\\
H       9.388972   26.652695    5.766540	\\
H      10.676751   27.911245    5.802432	\\
C      13.005089   25.665510    7.012250	\\
C      13.513339   26.561176    6.054061	\\
H      12.985913   27.517018    5.868582	\\
C      14.713268   26.281899    5.401472	\\
H      15.114912   27.003092    4.663350	\\
C      15.418169   25.105071    5.685767	\\
H      16.370640   24.891407    5.162735	\\
C      14.920698   24.193405    6.616684	\\
H      15.436848   23.237961    6.848087	\\
C      13.721749   24.467567    7.279225	\\
N      11.478065   20.440099   11.180512	\\
N      10.003855   22.178942   12.333361	\\
N       8.775266   23.494498   10.565036	\\
C      12.408593   19.997141    9.045486	\\
C      12.918357   19.125851    8.079778	\\
H      13.195354   19.548705    7.091340	\\
C      13.081014   17.774955    8.386179	\\
H      13.483453   17.082870    7.621038	\\
C      12.750828   17.293427    9.659596	\\
H      12.879629   16.219411    9.896550	\\
C      12.245944   18.156490   10.630929	\\
H      11.933575   17.765155   11.618521	\\
C      12.055948   19.518712   10.336676	\\
C      11.666559   20.505982   12.473250	\\
C      12.631942   19.682775   13.243120	\\
H      13.452680   19.331991   12.571416	\\
H      13.078346   20.279745   14.074642	\\
H      12.142121   18.787278   13.701064	\\
C      10.810985   21.480729   13.153574	\\
C      10.777720   21.687779   14.537103	\\
H      11.483502   21.153446   15.201722	\\
C       9.816902   22.561530   15.054735	\\
H       9.761617   22.737795   16.146542	\\
C       8.914799   23.190749   14.196371	\\
H       8.101547   23.821929   14.604057	\\
C       9.058508   23.008922   12.809644	\\
C       8.310229   23.711196   11.775950	\\
C       7.116812   24.523390   12.119889	\\
H       6.177053   23.930361   11.983475	\\
H       7.137163   24.854052   13.184148	\\
H       7.048056   25.426713   11.463985	\\
C       8.294030   23.971264    9.377957	\\
C       6.985741   24.402518    9.078548	\\
H       6.208726   24.395808    9.864623	\\
C       6.645107   24.768102    7.776819	\\
H       5.611433   25.100776    7.560535	\\
C       7.584354   24.666278    6.741623	\\
H       7.286273   24.909318    5.702311	\\
C       8.882809   24.228377    7.004842	\\
H       9.636329   24.102363    6.198341	\\
C       9.255876   23.909629    8.319972	\\
O      13.629165   25.693699   10.335326	\\
O      15.054352   23.949492   10.419150	\\
C      14.579768   25.047346   10.878908	\\
C      15.148944   25.577028   12.154026	\\
H      16.102522   25.070021   12.411438	\\
H      15.283882   26.679496   12.085672	\\
H      14.405273   25.377946   12.971604	\\
O       9.240174   20.965235    9.109180	\\
O       9.939535   21.279195    6.990173	\\
C       9.017738   21.132906    7.868470	\\
C       7.596402   21.202296    7.413526	\\
H       7.528188   21.226430    6.305692	\\
H       7.019675   20.346110    7.830899	\\
H       7.143116   22.142987    7.825555	\\
O      14.587882   20.910049   10.688605	\\
O      15.351206   21.082424    8.631351	\\
C      15.226349   20.400428    9.687313	\\
C      15.757963   19.019043    9.815996	\\
H      16.581455   19.010146   10.568026	\\
H      14.940826   18.348605   10.182828	\\
H      16.133811   18.655074    8.836485	\\
O      12.967978   21.908972    5.586999	\\
O      11.521639   23.566847    5.647444	\\
C      12.523106   22.978168    5.081699	\\
C      13.126088   23.607245    3.878303	\\
H      14.073814   23.096458    3.606186	\\
H      12.402614   23.547213    3.031790	\\
H      13.323867   24.686749    4.095232	\\

\bibliography{main}
\bibliographystyle{rsc}